\documentclass[12pt,preprint]{aastex}

\newcommand{\muh}{$\mu$ Her }

\slugcomment{ApJ, accepted}

\shorttitle{Apodizing Phase Plate Imaging}
\shortauthors{Kenworthy et al.}

\begin{document}


\title{First On-Sky High Contrast Imaging \\with an Apodizing Phase
Plate \footnote{Observations reported here were obtained at the MMT
Observatory, a joint facility of the University of Arizona and the
Smithsonian Institution.}}


\author{Matthew A. Kenworthy, Johanan L. Codona, Philip M. Hinz,\\
J. Roger P. Angel, Ari Heinze, Suresh Sivanandam}
\affil{Steward Observatory, 933 North Cherry Avenue, Tucson, AZ 85721}
\email{mkenworthy@as.arizona.edu}


\begin{abstract} 

We present the first astronomical observations obtained with an
Apodizing Phase Plate (APP). The plate is designed to suppress the
stellar diffraction pattern by 5 magnitudes from $2-9\lambda/D$ over a
$180\arcdeg$ region. Stellar images were obtained in the $M'$ band
$(\lambda_c=4.85\mu m)$ at the MMTO 6.5m telescope, with adaptive
wavefront correction made with a deformable secondary mirror designed
for low thermal background observations.  The measured PSF shows a halo
intensity of 0.1\% of the stellar peak at $2\lambda/D$ (0.36 arcsec),
tapering off as $r^{-5/3}$ out to radius $9\lambda/D$.  Such a profile
is consistent with residual errors predicted for servo lag in the AO
system.

We project a $5\sigma$ contrast limit, set by residual atmospheric
fluctuations, of 10.2 magnitudes at 0.36 arcsec separation for a one hour
exposure. This can be realised if static and quasi-static aberrations
are removed by differential imaging, and is close to the sensitivity
level set by thermal background photon noise for target stars with
$M'>3$.  The advantage of using the phase plate is the removal of
speckle noise caused by the residuals in the diffraction pattern that
remain after PSF subtraction. The APP gives higher sensitivity over the
range $2-5\lambda/D$ compared to direct imaging techniques.

\end{abstract}





\section{Introduction}

The direct imaging of extrasolar planets (ESPs) presents a daunting
technical challenge. The small angular separation from the host star
places the planet within a diffracted starlight halo many decades
brighter than the planet itself. A coronagraph suppresses this
diffraction whilst preserving the angular resolution and flux of the
planet through to the final imaging camera. There are many coronagraphic
designs and they all represent the fundamental trade-off between
throughput, angular resolution and diffraction suppression as a function
of position in the final image.

The prototypical coronagraph \citep{Lyo39} is an imaging camera
consisting of optics that form two intermediate focal planes. The first
is an image plane where an occulting mask is used to remove the Airy
disk and diffraction rings of the central star. A sharp edged mask in
the image plane introduces high spatial frequency components in the
wavefront, which are masked out in the subsequent pupil plane with an
undersized pupil stop (the Lyot stop). The original design is not
optimal for ESP detection, and results in reduced throughput and angular
resolution.

Encouraged by the discovery of nearly two hundred ESPs through radial
velocity searches around nearby stars \citep{But06}, the potential of
directly imaging terrestrial sized planets has fueled research in a wide
range of coronagraphic designs, a large selection of which are reviewed,
analyzed and compared in \cite{Guy06}.  In that paper, they concentrate
on designs suitable for space-based telescopes consisting of an
unobstructed circular aperture, which produce contrast ratios in the
regime of $10^{-10}$. Most of the designs considered modify the incoming
wavefront by placing apodizing masks in the image plane and/or in the
pupil plane of the coronagraph. The designs can be further subdivided by
whether they modify the amplitude or the phase of the wavefront, or some
combination of the two.

Most of the image plane apodizing methods are sensitive to
tip-tilt errors and the finite size of the stellar disk \citep{Guy06},
allowing light from the primary star to add an extra source of noise and
reduce the achievable contrast\citep{Kuc02}.  Recent theoretical studies
have indicated that eighth-order band limited masks, which are designed
to be insensitive to small amounts of low order aberrations
\citep{Kuc05,Sha05}, show promise for providing deep suppression over
small opening angles and with theoretical throughputs of up to 50\% and
IWAs (inner working angles) of $4\lambda/D$.  Laboratory experiments on
manufacturing process limited masks gave a measurement of 15\%
throughput for a designed transmission of 20\% \citep{Cre06}.

Alternative methods for suppressing diffraction rely on variable
transmission masks in the pupil plane, either as binary masks
\citep{Kas03,Van03} or as graded (apodized) transmission masks
\citep{Nis01,Aim05}. Pupil apodizing methods generate field invariant
PSFs and so do not suffer from the tight alignment tolerances required
by image plane masks, and their suppression is not affected by the
angular size of the target star.

\cite{Guy06} considered a phase apodization algorithm from
\cite{Yan04}, which uses a square pupil and provides deep suppression
over a 90 degree opening angle with a 60\% loss of light from the
central PSF core. Our independently developed phase apodization
algorithm can generate PSFs for more generalized pupil geometries
(including pupils with secondary obscurations) over an opening angle of
180 degrees. For the phase plate in this paper, the FWHM of the Airy
core increases by 10\% and produces a considerably smoothed zone of
suppressed diffraction from 5 to 6 decades below that of the core, at a
modest cost of light (29\%) from the central Airy core of the imaged
object and an IWA of $2.2\lambda/D$.

The sensitivity of close-in searches is increased by the subtraction of
the PSF of the telescope, determined either by an analytical model,
images of a nearby reference star of similar brightness, or by using
the rotation of the target star on the night sky to construct a model
PSF with the putative companion signal removed by median combining the
individual exposures. In the ideal high signal-to-noise limit, the PSF
images subtract cleanly, leaving an image limited by photon noise
following the diffraction pattern.  In reality, PSF subtracted images
show considerable small scale structure, and these systematic
aberrations (thought to be caused by optical flexure within the
telescope and camera system) limit the detection sensitivity of faint
companions at small angular separations.

The 6.5m MMT AO system \citep{Wil03,Bru99} is the first realization of a
large aperture telescope coupled with a deformable secondary mirror that
delivers an AO-corrected $f/15$ beam. At longer wavelengths $(>2.5\mu
m)$, the MMT AO system is uniquely sensitive because of the lower level
of background light it emits at infrared wavelengths compared to
conventional AO systems on larger telescopes \citep{Llo00} -- typically
twice as efficient as the Keck telescopes at $5\mu m$.  Furthermore, the
mid-infrared wavelengths are well suited to the detection of hot
extrasolar planets due to their enhanced thermal emission detectable
through the Earth's atmospheric transmission window at 5 microns
\citep{Bur03,Bur04}. Where thermal telescope emission is the dominant
source of background photons, a reduction in emissivity corresponds to a
direct reduction in exposure time to reach a particular sensitivity.
Using measurements of sky frame fluxes at two differing airmasses, the
emissivity of the telescope, AO system and Clio camera is estimated to
be 10\% \citep{Siv06}. This is a significant improvement over the
25-50\% of other conventional mid-IR AO systems.

In this paper we present a new method to suppress diffraction in an
astronomical image: the Apodizing Phase Plate (APP).  In Section
\ref{ppd} we present the basic theory behind the optic, and Section
\ref{obs} describes the first on-sky APP observations and the data
reduction.  Section \ref{expect} compares the expected point spread
function using the APP with the MMT AO system, and the calculated
sensitivity and limits to the achieved contrast ratios are in Section
\ref{compare}.  Section \ref{muher} presents an initial scientific
observation using the APP and we discuss the implications for future
observations in Section \ref{concl}.

\section{Phase Plate Design and Manufacture\label{ppd}}

\newcommand{\kvec}{\vec{\kappa}}
\newcommand{\xvec}{\vec{x}}
\newcommand{\thvec}{\vec{\theta}}

The starlight halo against which detection of faint companions are to be
made consists of a deterministic diffraction pattern (or point spread
function, PSF), and a speckled halo caused mostly by residual phase
errors left over by the AO system, and to a lesser degree by
scintillation over the pupil.  There are also inevitably going to be
optical imperfections in the optics, e.g. small phase aberrations and
transmission variations that will lead to faint static and semi-static
speckles in the halo.  For the MMT, these have not yet been seen or
characterized.  The scintillation halo is expected to be fainter than
the AO-residual phase halo, and at high Strehl, the diffraction pattern
is the brightest.  Reducing the diffraction halo is the first step in
improving the S/N of a faint planet near the star, since it reduces the
``noise'' against which the detection is made.  Preserving the light in
the diffraction core is also important since it affects the ``signal''
coming from the planet.  The phase plate is intended to suppress the
deterministic diffraction pattern to a level below that of the next
limiting component of the halo (i.e.~the residual AO phase errors),
while doing the least harm to the light in the diffraction core.

The phase plate was designed using the approach outlined as ``Method I''
in \cite{cod04}. This method has recently been validated in a
closed-loop laboratory experiment \citep{Put06}. The underlying
principle is to create an ``anti-speckle'' for each ``speckle'' or
diffraction structure within some region of interest (ROI) within the
PSF of the star. (An ``anti-speckle'' is an intentionally-constructed
region of the halo which matches an unwanted speckles' location and
amplitude, but is 180 degrees out of phase.  The anti-speckle is
intended to cancel a given speckle).  We accomplish this by introducing
a weak sinusoidal phase ``ripple'' across the pupil which acts as a weak
diffraction grating, causing a small amount of the starlight core to be
diffracted into a new halo speckle. Strictly speaking, a line of
speckles are formed corresponding to the higher diffraction orders of a
grating, but in the limit of a small sinusoidal ripples, the higher
orders can be neglected.  By using a phase ripple with spatial frequency
$\kvec=2\pi\thvec/\lambda$, we can create a speckle at an angle
$\vec{\theta}$ relative to the star.  For a small phase ripple, the
amplitude of the created speckle is proportional to the ripple
amplitude, and the speckle phase relative to the core is linearly
related to the ripple phase. Since the phase ripples are real (i.e.~only
affect phase and not amplitude), each created speckle has an
anti-Hermitian (antisymmetric real, symmetric imaginary) counterpart on
the other side of the star. This is the opposite symmetry of diffraction
patterns, and therefore it is only possible to suppress diffraction over
at most one half of the region around the star. The region on the other
side of the star is reinforced, adding energy to the diffraction pattern
there.

By linearly superimposing a set of such ripples, we are able to reduce
the halo over the region of interest (ROI). Since this algorithm does not take into account
the detailed shape of the created speckles, and the process of creating
them alters the initial halo in a nonlinear fashion, this approach will
only suppress, not cancel, the halo in a single step.  However, by
iteration, we can robustly find our way to a solution which nearly
cancels the halo over the ROI. For small iteration steps, the formula
for the $n+1$ iteration of the phase is \citep{Cod06}

\begin{equation} \varphi_{n+1}(\xvec)=\varphi_{n}(\xvec)-\Im\left\{e^{-i\varphi_{n}(\xvec)}\int_{\Pi}d^{2}x'e^{i\varphi_{n}(\xvec')}\mathcal{H}(\xvec -\xvec')\right\} ,\label{eq:phaseUpdateSimplified} 
\end{equation}

where $\Pi(\xvec)$ describes the pupil, $\mathcal{H}(\xvec)= \int_{ROI}d^{2}\kappa\,\exp\left(i\kvec\cdot\xvec\right)/(2\pi)^{2}$ is the spatial filter corresponding to the dark region of interest in
the focal plane, and $\Im$ takes the imaginary part of the result. Equation \ref{eq:phaseUpdateSimplified}
is derived by making the halo within the ROI dark without regard for the effect on the PSF core or the halo outside of the ROI. While this approach worked adequately in this case, it is sub-optimal in terms of Strehl ratio. The theory has now been extended to include the maximum preservation of power in the diffraction core. The extended theory will be presented in a later paper.

Suppressing diffraction by introducing aberrations is inevitably chromatic.
The practical effect is that the dark zone in the halo begins to brighten
with increasing bandwidth. However, so long as the combined halo remains
at least an order of magnitude below the residual AO speckle halo,
it won't significantly limit detection. Since the mismatch between
the complex antihalo and the halo grows linearly for small variations
in wavelength, the intensity of the halo floor brightens like
$(\lambda-\lambda_{0})^{2}$
where $\lambda_{0}$ is the wavelength where the eq.
\ref{eq:phaseUpdateSimplified}
iteration was performed \citep{Cod06}. 
Integrating the growing halo over the entire
detector band shows that the total halo background grows as the cube
of the bandwidth. This leaves us a reasonable bandwidth with which
to work, effectively controlled by the Strehl ratio achieved by the
AO system. In general, depending on the science goal, working with
a wider bandwidth is better, since it collects more photons and gives
better speckle noise averaging. Therefore, it is best to use as wide
of a bandpass as possible until the diffraction halo starts to become
significant. 

The mean residual AO speckle halo is unaffected by
the phase plate and remains at the same level, while the Strehl ratio
of the phase plate drops the PSFs of the star and planet into the
noise. For this reason, it is better not to let the iteration run
to something like convergence, but rather to stop it as soon as the
solution becomes ``good enough.'' Since the Strehl ratio starts
at unity and drops monotonically as the iteration is performed, it
is easy to place a threshold and stop the calculation when the desired
level is reached.

Two APPs were manufactured to our computed specifications by {\em II-VI
Incorporated}, both using zinc selenide (ZnSe, $n(4.68\mu m)=2.43$) as the
transmitting substrate for the phase pattern. The wavefront surface was
generated using a diamond tool whose pitch depth was synchronized with
the rotation angle of the machine lathe holding the APP. The first APP
used a phosphor bronze plate aligned with the phase plate pattern to
mask the telescope pupil and secondary mirror support structures.
Initial tests with this first APP validated the APP method, but the
plate has no anti-reflection coatings on either side, decreasing the
transmission efficiency of the optic and adding a scattered halo of
light from the two internal reflections of the optic. The success of
this plate demonstrated that pupils of arbitrary geometry, such as
segmented mirrors on space telescopes, can be accommodated into the APP
algorithm and produce a nulled region near the target of interest. We
carried out scientific observations (detailed in Section \ref{muher})
with this engineering version, and we were encouraged with the results
to make a second, science grade APP.

In the case of thermal imaging at the MMT, we also concluded that the
thermal emission from the secondary support vanes does not significantly
contribute to the noise background of our observations.  Removing the
secondary support masks from the APP design has two advantages - (i) the
APP does not have to rotate to keep aligned with the secondary support
vanes, resulting in cleaner background subtraction at thermal
wavelengths, and (ii) the APP surface does not contain high spatial
frequency regions required to deal with the presence of the secondary
support vanes, resulting in a wavefront that is less sensitive to
manufacturing errors in the diamond turning process.

\begin{figure}[ht!]
\begin{center}
\includegraphics[angle=270,scale=1.1]{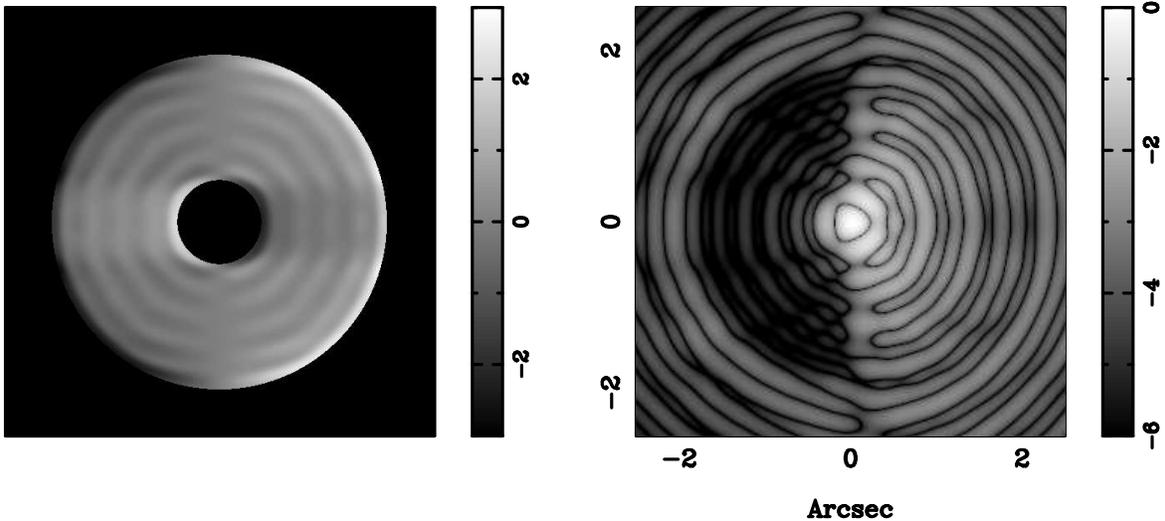}
\end{center}

\caption{Phase map of the APP design and its resultant theoretical point
spread function.  The left hand image shows the phase plate design cut
into the zinc selenide plate, with the scale bar on the right showing
height in microns. The secondary obscuration is deliberately over sized
to allow for rapid in-situ alignment. Shown on the right hand panel is
the calculated PSF logarithmically scaled over 6 decades normalized to
the peak intensity. The pixel scale on the right hand panel is for 5
micron imaging at the MMTO 6.5m telescope.  \label{fig:theory}}

\end{figure}

The science grade APP is manufactured as a single optic that contains
the phase plate pattern, with a gold layer coating that blocks light
outside the telescope pupil and within the secondary mirror's shadow. A
broadband coating of magnesium flouride with a measured reflectivity of
less than 0.5\% (from 3.2 to 5.5$\mu m$) provides high optical
throughput and minimizes ghost image reflections. Figure
\ref{fig:theory} shows the design of the APP and the resultant PSF. The
plate is designed to suppress diffraction from $2-9\lambda/D$ in the
focal plane of the camera over a $180\arcdeg$ region, forming a D-shaped
region of suppression next to the target star.


The surface accuracy of the APP was measured using a Zygo optical
interferometer at the manufacturers optical laboratory. The mean
departure from the model surface is measured to be 70 nm r.m.s. Using
the Marechal approximation for estimating Strehl ratio at the operating
wavelength of 5 microns, this corresponds to approximately 0.7\% of the
incident flux scattered out of the transmitted beam due to
microroughness in the APP.

The inner radius is set by signal to noise considerations for faint
companions. Suppression of the Airy pattern comes at a cost of energy
from the core, and for this design we chose an APP core signal (the flux
enclosed within the first dark Airy ring) that was 69.4\% of the direct
imaging PSF core signal. The outer working radius of the nulled region
was set to match the control radius of the MMT AO system.  Beyond the
first Airy ring, the suppressed halo is designed to be 5 ($3\lambda/D$)
to 6 ($6\lambda/D$) magnitudes fainter than the direct imaging PSF at a
similiar radius, although we did not actually validate that claim by
measurement.  The phase plate theory is capable of creating much deeper
suppressions, although in practice the limit will be set by the
microroughness of the manufacturing process.  One notable advantage of
using the APP at the pupil plane of the telescope is that the
suppression is independent of tip-tilt and pointing errors in the
telescope, which allows easy beam switching on the infra-red detector.

Although it is possible to use the deformable mirror of an AO system to
produce the APP wavefront, it is more practical to implement a single
optic that sits downstream of the AO control loop.  The two main reasons
for this are: (i) the prescribed wavefront has many high spatial
frequency features that require a high degree of actuator control to
hold in place, and (ii) the wavefront sensor camera will see the
wavefront superimposed on the atmospheric turbulence, potentially
saturating the sensor. Keeping the APP out of the control loop allows
the deformable mirror to be used in the removal of low-level residual
static aberrations in the ROI of the APP, and to test any candidate
companions for coherence with the light from the primary star
\citep{Ken06}.

\section{Observations\label{obs}}

Clio is an imaging camera \citep{Fre04,Siv06} designed for obtaining
high spatial resolution images with optimum efficiency at L' and M band.
It is optimized for imaging extrasolar planets at wavelengths where they
are expected to have fluxes significantly in excess of that of a similar
temperature blackbody \citep{Bar03,Bur03}.  The plate scale of Clio is
determined to be $0.048574\pm 0.000090$ arcsec pixel$^{-1}$ from
observations of two binary systems (HD 100831 and HD 115404) in April
2006.  The plate scale error is dominated by uncertainty in the true
separation of these binaries. Position angle (PA) calibration was also
determined from the two double stars mentioned above, and the accuracy
is 0.20 degrees, again limited by the uncertainty in the true position
angles of the double stars.

The APP was placed in the Clio camera at the intermediate pupil plane
and mechanically positioned into the telescope beam using a motorized
filter wheel. The camera has a pupil imaging mode which allows quick and
easy alignment of the APP with the telescope pupil. To go from direct
imaging to APP imaging takes less than five minutes, which is in strong
contrast with systems involving an occulting mask in an image plane that
require precise alignment for each new target observation.

The APP was used on M' ($\lambda_c=4.85 \mu m$) observations of HD
213179 (V=5.778 SpT=K2II) taken on 2006 July 09 12:10 UT. Extrapolating
to M' magnitudes using color tables in \cite{Cox00} this star has an
approximate magnitude $M'=3.1$, which is the typical brightness of stars
looked at in extrasolar planet surveys carried out with the Clio camera.
The data were taken at an airmass of $1.06$ in conditions of variable
seeing, typically $0.6-1.0$ arcsec at 500nm as estimated by wavefront
sensor camera residuals in the AO system. We obtained in a set of 16
5-second exposures on target and 16 off target for a total of 80 seconds
on-source integration.

\begin{figure}[ht!]
\begin{center}
\includegraphics[angle=270,scale=0.80]{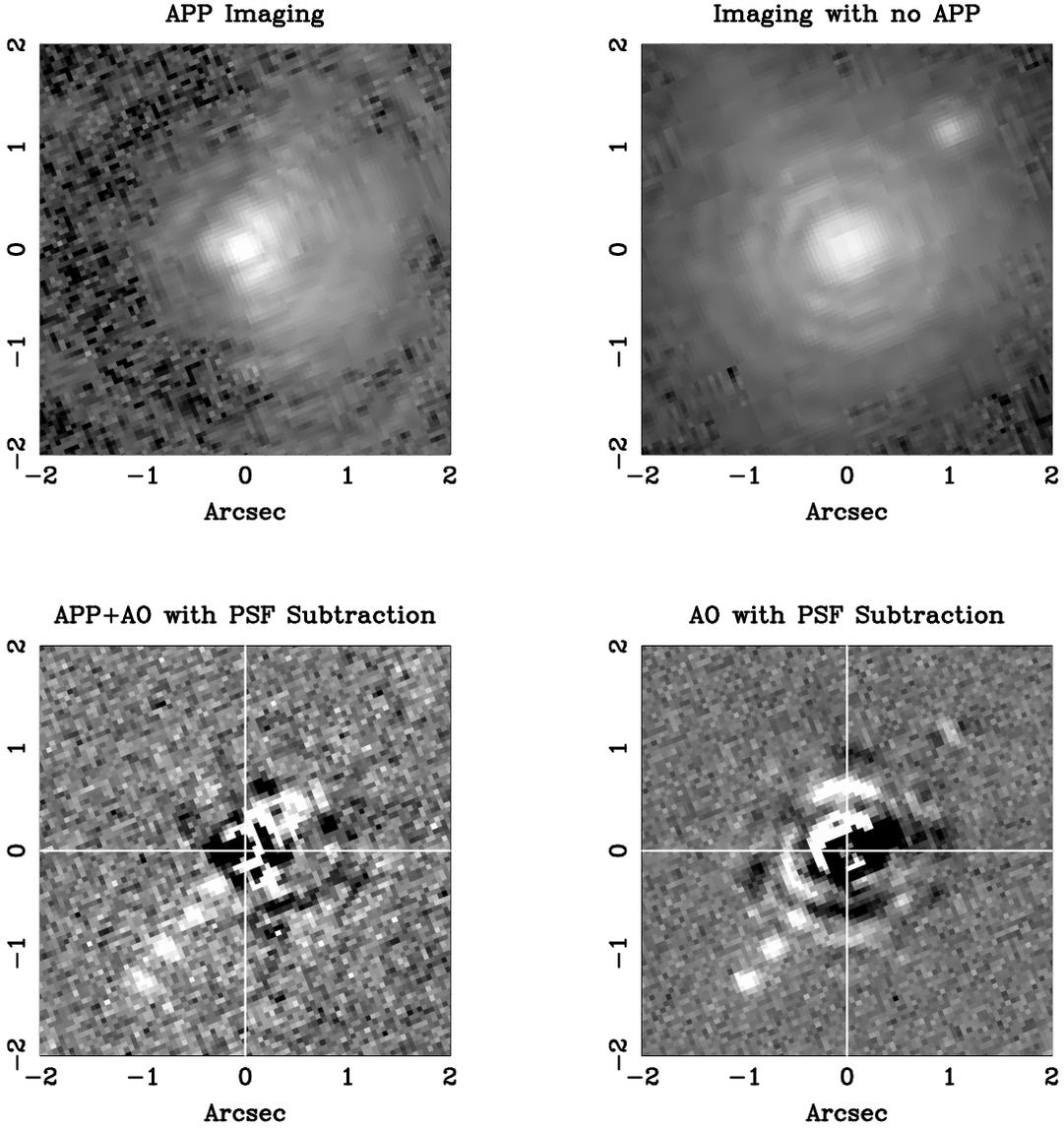}
\end{center}

\caption{Comparison of AO imaging with and without the APP. The upper
row shows AO+APP imaging of a single star (HD 213179) and AO imaging of a
binary system ($\mu Her A$), with the flux and sky background of both
images scaled to represent identical exposure times of stars with
identical magnitudes.  Both figures are logarithmically scaled so that
the APP peak intensity has 69\% of the no plate imaging intensity,
consistent with the APP design.  The lower two panels show the detection
sensitivity of the two PSFs under PSF subtraction.  Fake sources with a
magnitude 7.5 fainter than the star have been added at 0.4 arcsecond
increments $(2\lambda/D)$ from 0.4 to 1.6 arcseconds. The APP image
intensity is scaled so that the fake source has the same signal in both
PSF subtracted images, showing that even with the lower signal to noise
the sources are more easily identified with the smoother, suppressed
background and are not lost in the systematic Airy ring PSF
subtraction.\label{fig:psfsub}}

\end{figure}

Approximately 1\% of the pixels on the detector show anomalous
behavior,  including high dark current, high or low sensitivity, and
time-varying responsitivity. Out of the methods we explored, we found
the most robust way to select anomalous pixels was by looking at the
sky-illuminated pixels in the beamswitched images. For a set of
exposures making up a single beam, the variance of the sky illuminated
pixels is calculated. Any pixels that have significantly high or low
variance are marked as bad pixels.  Combining the bad pixel masks from
the two beam pointings gives a bad pixel mask that is applied to all of
the unshifted science frames from the camera, interpolating over the bad
pixels using adjacent good values. A blank region of the array is then
used to sample a residual offset caused by the time-varying sky
background, which is then subtracted from the image. The cleaned images
are co-aligned using the IRAF imalign routine, which performs a
cross-correlation in a box of $5\times5$ pixels centered on the
unsaturated target star to locate the centroid of the PSFs. Resampling
of the images so that the centroids of the PSFs are coincident is done
using a cubic polynomial interpolant. These are combined with rejection
of the most extreme pixel values to form the final image. The cleaned
and background subtracted images were coadded to yield the summed image
in the upper left panel of Figure \ref{fig:psfsub}.  The PSF with the
APP can be compared with that for the direct telescope PSF in the same
figure. The Airy rings are strongly suppressed in the APP image.


The utility of the APP becomes apparent when compared to direct imaging
and PSF subtracted images. Figure \ref{fig:psfsub} shows the telescope PSF
comparing direct imaging with APP imaging. The Airy rings are strongly
suppressed in the APP image, with only the time smoothed servo lag halo
visible within 0.5 arcseconds of the target star. This is very much in
evidence in the PSF subtracted images in the lower row. Here, four
fake sources have been added to half the dataset and the other half of
the data are used to subtract off the PSF. The background noise has
been scaled in both the direct and APP imaging to represent identical
on-source integrations of 40 seconds through the M' filter. The APP is
background noise limited up to $2\lambda/D$ of the target star,
whilst in the direct imaging case the closest fake source is confusion
noise limited by the Airy ring residuals.

The ``Measured Light Profile'' line in the upper panel of Figure
\ref{fig:radplots} represents the measured light profile of the Clio APP
image for a total on-sky integration of 80 seconds, normalized to the
mean flux in the Airy disk of the star. The flux at a given radius is
calculated by azimuthally averaging over a 150 degree wedge centered on
the target star in the middle of the ROI. This curve can be compared
with the dotted line labeled ``Turbulence-free Model'', which
represents the theoretical APP PSF in the limit of perfect AO correction
over the M' band. Comparing the measured light profile and
turbulence-free model in the upper panel of Figure \ref{fig:radplots}, we see
that the measured starlight in the region of the suppressed halo is 4 to
5 magnitudes brighter than the theoretical profile.  This brightening is
due to the residual aberration of the AO corrected wavefront, as
discussed in the next section.

\section{PSF modeled for the AO-corrected wavefront\label{expect}}

The Strehl ratio for our observations is typically around 90\% (see next
section for calculation).  The ultimate limit to high-contrast imaging
is from low-order (a few cycles/aperture) wavefront errors which cause
speckles within a few $\lambda/D$ of the star.  The diffraction
suppression provided by the APP allows us for the first time to directly
measure the residual speckle intensities resulting from low-order (a few
cycles/aperture) wavefront errors, allowing us to estimate the low-order
performance of the AO system.

\begin{figure}[ht!]
\begin{center}
\includegraphics[angle=0,scale=0.90]{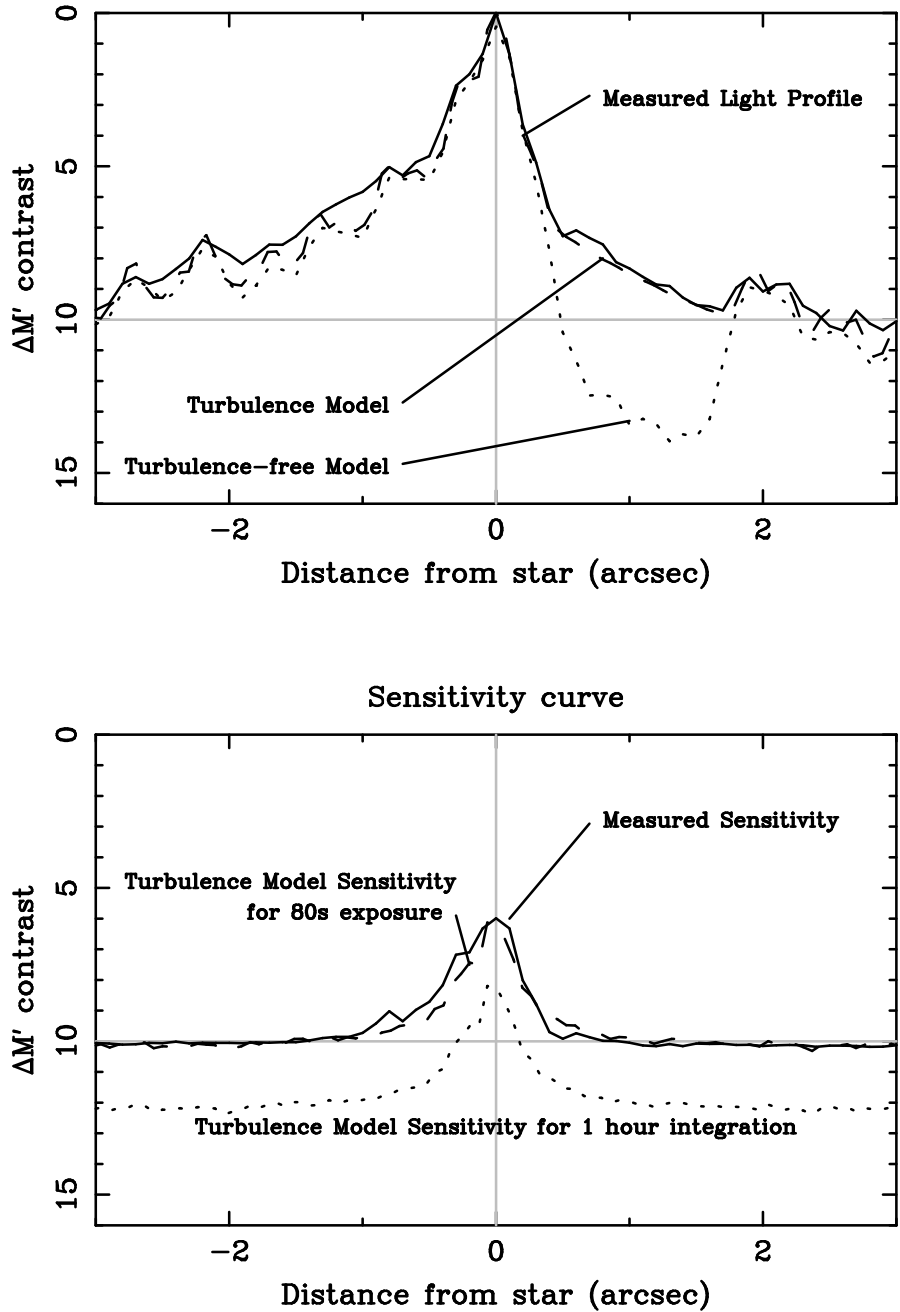}
\end{center}

\caption{Measured light profile and contrast ratio of the APP. The upper
panel shows the azimuthally averaged light profile for a 150 degree
wedge centered on the PSF for different cases described in the text, and
the lower curve shows the $1\sigma$ noise level, representing the
sensitivity curve for the upper plot. The procedure for generating these
curves is discussed in Sections \ref{expect} \& \ref{compare}.  \label{fig:radplots}}

\end{figure}

Using the APP has an effect on both the signal and the noise portions of
the detection sensitivity.  The APP is designed to improve detection
sensitivity by suppressing the diffraction halo, but at the cost of
light taken by the plate from the PSF core.  Furthermore, other sources
of background light will become the sensitivity constraint once the
diffraction halo has been sufficiently suppressed, making deep halo
suppression and its concomitant loss of Strehl ratio a liability.  For a
bright star, the next constraint in the sensitivity is the residual AO
speckle halo, while for a fainter star the constraint might be photon
noise from the thermal sky background.  The azimuthally averaged light
profile for the measured Clio APP data is shown as the solid line in the
upper panel of Figure \ref{fig:radplots}.

Over the suppressed-diffraction region of the halo, the residual
scattered halo is consistent with an angular dependence of
$\theta^{-5/3}$, which can be explained with a time-lag in the applied
AO correction \citep{Angel2003}.  This halo results when the wind
advects the aberration-causing turbulence past the telescope aperture
and there is a processing lag, $\tau_{lag}$, between the wavefront
measurement and the application of the corresponding correction.  Since
the processing lag and the wind velocity multiply to give the shift
between the actual wavefront and the AO-corrected version, the lag error
can be parameterized by the distance $d_{lag}$ which is the length of
$\vec{v}\tau_{lag}$ where $\vec{v}$ is the mean wind velocity.

\clearpage

\clearpage

To understand and quantify the AO halo in conjunction with the APP, we
simulated our AO system with the APP phase profile to model the combined
PSF. As described in \cite{Angel2003}, a single Taylor wind flow results
in a halo of speckles that are not evenly distributed with azimuth angle
$\phi$, but instead form a two-lobed distribution (proportional to
cos$^2\phi$) aligned with the wind direction projected on the sky.
Since this was not an obvious feature of our data, we modeled the wind
as two superimposed turbulence layers oriented $90\arcdeg$ from each
other, resulting in a reasonably isotropic speckle halo.  The two
simulated winds were given the same speeds, but with independent
Kolmogorov phase screens \citep{Har03}.  The instantaneous residual
phase error was computed by shifting and summing the two wind layer
phase screens to the positions appropriate to time $t$, and then
subtracting a low-pass-filtered version of the combined phase
appropriate to time $t-\tau_{lag}$.  The resulting residual wavefront
was then passed through a Fourier optics model of the MMT, including the
APP phase, resulting in an instantaneous PSF of the telescope.  The
phase screens were computed using a Fourier synthesis method on a
$2048\times2048$ point grid, modeling a periodic 32m square region
characterized by Kolmogorov turbulence with a characteristic Fried
length $r_0$.  The evolving PSF was computed for 256 steps across the
screen, shifting the array to simulate the wind evolution over a time of
$32/v$ seconds. Each frame represents an instantaneous realization of
the atmosphere. By averaging together all the 256 frames, the resultant
PSF represents a telescope integration that covers a 32m patch of
atmosphere moving across the telescope pupil.  The 5 second exposures
from the Clio camera fix the exposure time, and so a PSF for a given
length of atmosphere $v \tau_{exposure}$ (and therefore a given wind
velocity) can be constructed by adding up a smaller subset of the 256
frames.  Since a smaller number of exposures sums over fewer speckles,
the variance in the individual exposures will be higher.

By varying the wind lag distance and the Fried length in the AO models,
we produced different modeled contrast ratio curves, the best fit of
which is shown as the ``Turbulence Model'' in the upper panel of Figure
\ref{fig:radplots}.  The best fit parameters are $r_0=20$ cm at the
wavelength of $500$ nm and a wind lag distance of $19$ cm in each of the
two wind directions.

These parameters were then used to generate 16 simulated exposures,
processed in an identical manner to the real dataset. Both the 16 Clio
exposures and the 16 simulated exposures are averaged and the per pixel
variance of the individual exposures about the resultant mean was
calculated. This variance is divided by the number of images averaged
together and then square rooted. The azimuthal mean of this quantity is
then plotted as a function of radius over the image region identical to
that in the upper panel, and we refer to these as our ``Measured
Sensitivity Curve'' and ``Turbulence Model Sensitivity for 80 seconds
exposure'' in the lower panel of Figure \ref{fig:radplots}. These
represent the $1\sigma$ sensitivity limit of the APP in detecting a
faint companion using model PSF subtraction (see next section).

The width of the sensitivity curve is determined by the wind speed
during the 80 seconds of total integration. By varying the number of
summed frames from the total block of 256 frames, the best match to the
sensitivity curve was determined to be 192 frames, implying that
$(192/256)(32 \mathrm{m}) \approx \vec{v}\times 0.5 \mathrm{s}$, which
yields a mean wind velocity of $\vec{v}=4.8ms^{-1}$.  Recent
measurements of wind velocity as a function of altitude using SCIDAR
\citep{Pri04} show that the velocity of turbulent layers at altitudes of
10km can be considerably larger than the wind velocity measured at
ground level ($25ms^{-1}$ compared to $3ms^{-1}$ for observations at San
Pedro Martir in 2000), and this trend is consistent with our measured
ground layer wind velocity of $1.4-2.5ms^{-1}$.


\begin{figure}[ht!]
\begin{center}
\includegraphics[angle=270,scale=1.00]{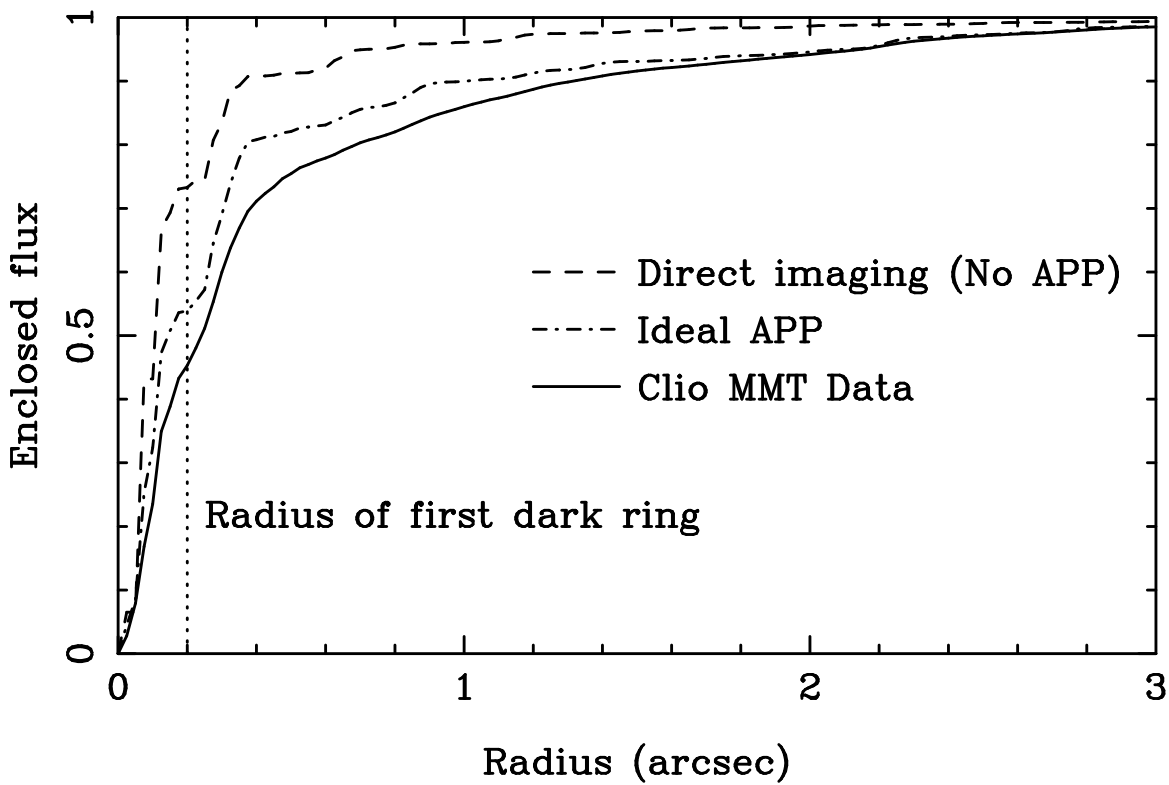}
\end{center}

\caption{The enclosed flux as a function of radius for
various PSFs.  This plot compares the direct imaging PSF with the
idealized APP PSF and the measured APP PSF. The radius of the first dark
ring indicates the flux enclosed within the central Airy disk.
\label{fig:ee}}

\end{figure}


To determine how well our AO system was performing with the APP at 5
microns, we wanted to determine what the Strehl ratio of our data were.
Figure \ref{fig:ee} shows the the fraction of the total flux enclosed
within a given radius for various PSFs. The energy within the radius
of the first dark ring indicates the fraction of energy in the image
core. The dotted line represents the PSF of the telescope pupil,
showing that the encircled energy of the core is 74\%, consistent with
a pupil with central obscuration. The dot-dashed curve
represents the modeled PSF of the telescope with an ideal APP, which
is 69.8\% of the encircled energy of the ideal PSF, or 52\%. The data
from Clio with the APP gives an encircled energy of 45\%. The ratio of
the encircled energy from the measured APP to the ideal APP gives an
estimate of how efficiently the AO system is working at keeping flux
within the core of the image (and hence, the flux of the planet in its
core) and this is approximately 86\%.

This encircled energy ratio is analogous to the Strehl ratio used in
quantifying the wavefront correction of AO systems, and this is
consistent with the Strehl ratio expected for the MMT AO system running
at 550Hz with a 56 Zernike mode level of correction. Using the Marechal
approximation we derive a mean wavefront error of 220nm. 

\section{APP Sensitivity Comparison\label{compare}}

An observation looking for point-like sources goes further by performing
PSF subtraction on the data. For scientific observations over many hours
of on-source exposure, the camera and phase plate are fixed relative to
the telescope during the set of exposures, so that the sky and
background light distribution within the whole optical system does not
systematically change. For our alt-az mounted telescope, the sky then
rotates as a function of hour angle.

We construct a master PSF by taking the median of the individual
exposures, which rejects the flux from faint compansions whilst building
a high signal to noise PSF that includes the static aberrations in the
system. This master PSF is subtracted off the individual exposures, and
each frame is rotated in software to align the sky orientation with the
image orientation. The images are added together to form a final
PSF-subtracted image.

The ``Measured Sensitivity'' and ``Turbulence Model Sensitivity'' in the
lower panel of Figure \ref{fig:radplots} closely match each other, and
so we conclude that the total integration of 80 seconds represents a
significant sampling of the turbulence power spectrum of the atmosphere.
If we assume that the sensitivity curve will increase as the square root
of the number of coadded exposures, we can estimate the expected
sensitivity for a one hour exposure, labeled as ``Turbulence Model
Sensitivity for 1 hour integration''. This sensitivity curve is
dominated by two components - one attributed to the photon shot noise of
the sky background which scales as $t^{0.5}$ \citep{Hin06} and is
present at large radii, and a residual speckle halo noise component. In
the limit of the residual speckle halo noise and sky background noise
statistics both reducing as $t^{0.5}$, the halo noise is fixed with
respect to the sky background noise, and the transition point for
background noise dominating over halo noise is at about 0.8 arcseconds
($4\lambda/D$). If the star is brighter than M'=3, then the residual
speckle halo noise will be dominant out to larger radii, and if the star
is fainter, the sky background noise will become dominant at smaller and
smaller radii.

Assuming that other static and quasi-static aberrations do not become
dominant in longer integrations, we expect to be able to detect sources
of $\delta M'=10.2$ with a $5\sigma$ confidence level in one hour at a
separation of $2\lambda/D=0.36"$ with the estimates of sensitivity made
with our AO simulations.  Our next investigation will be to obtain a
much longer exposures with the APP and determine what sources of noise
appear as the sky background noise is reduced, and to develop a suitable
theory to characterize them.


\begin{figure}[ht!]
\begin{center}
\includegraphics[angle=270,scale=1.00]{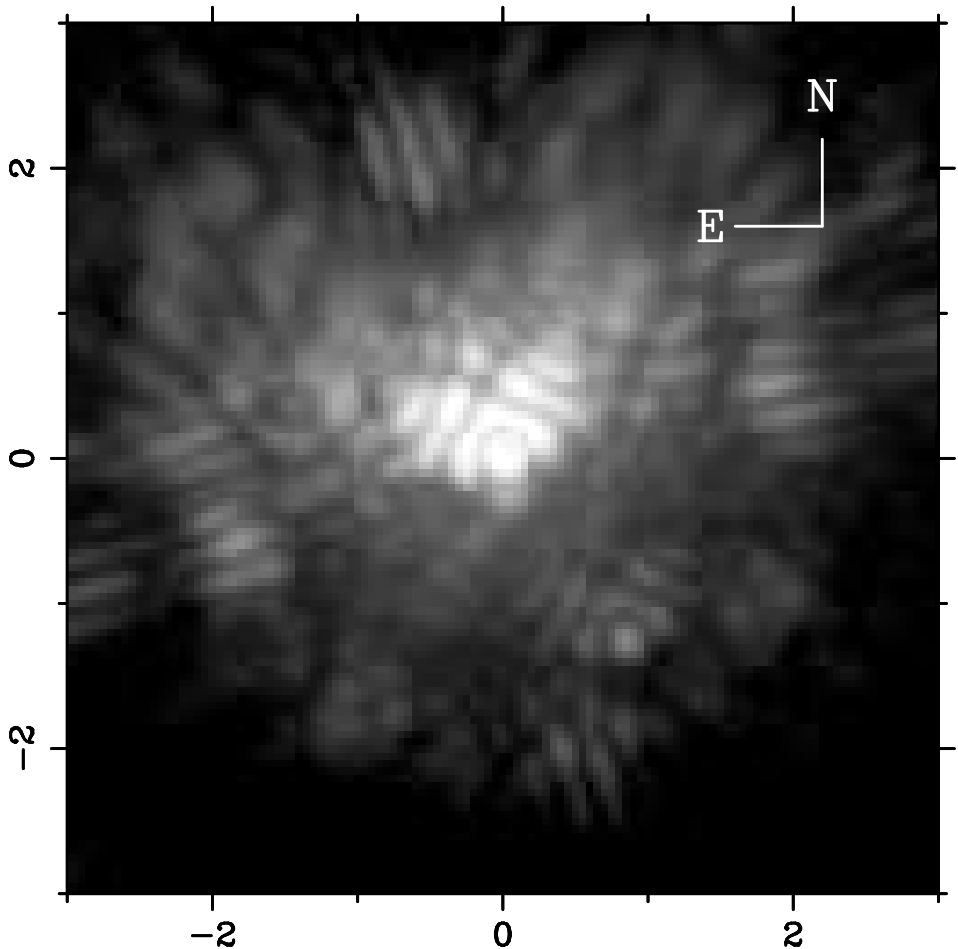}
\end{center}

\caption{Image of the \muh A system taken with an earlier
prototype APP.  Note how the secondary companion (to the lower right)
has the same PSF as the primary star.\label{fig:muher}}

\end{figure}


\section{The \muh A System\label{muher}}

In order to demonstrate that the principle of the APP was valid, we
observed a recently nearby binary system, with the purpose of confirming
the nature of the fainter component as determined by \cite{Deb02} and to
obtain astrometric measurements for determining the dynamical mass of
the system.

The star \muh A was observed with the earlier prototype of the APP at
2006 April 13 UT 12:48 using the Clio camera at the $f/15$ Cassegrain
focus of the MMTO 6.5m telescope on Mount Hopkins, with the AO system
operating in closed loop on the target star. The data were observed in
the M' filter ($\lambda_{c}=4.67 \mu m, \lambda_{fwhm} = 0.25$).

Our astrometric and photometric observations of the \muh A system are
listed in Table \ref{tbl-2} with the image shown in Figure
\ref{fig:muher}. Although we performed relative photometry in both M and
M', our images are saturated (and thus non-linear) by a few percent in
the M band images, and so that relative photometry is not reported here.
\cite{Kid04} give $J,H,K$ photometry for the \muh A system that used a
15 arcsecond aperture, including both the components. Our M' band
differential photometry, combined with the typical $(K-M=0.005)$ color
of \muh A and $m_{K}(Aa)=1.743$ results in an absolute magnitude of
$M_M'=7.13\pm0.05$. The resultant colors for \muh Ab are
$(H-K)=+0.6\pm0.12$ and $(K-M')=+1.17\pm0.12$, where errors are
predominantly from the H and K band magnitudes, $M_K=8.3\pm0.1$ and
$M_H=8.9\pm0.1$.

Examining the photometry of 3 Gyr M, L and T dwarfs presented in
\cite{Gol04,Leg02}, we find that the absolute magnitudes are indicative
of a mid-M dwarf, consistent with the $(H-K)$ color. Using the
polynomial fit in \cite{Gol04} for the spectral type as a function of
$M_{M'}$, \muh Ab is determined to be M4$\pm1$. Using the photometry
from \cite{Rei02} the $M_K$ implies a later spectral type, around M6-8.
Even assuming a later type star, the $(K-M')$ color is still a whole
magnitude too red for an M4 dwarf but similar to that for an early T
dwarf, a conclusion ruled out by the age and the absolute magnitude of
the object. Low resolution infra-red spectroscopy of this low mass
object, along with further astrometric observations over the next decade
will help determine its true nature.

\section{Conclusions and Future Plans\label{concl}}

We have demonstrated for the first time that an apodizing phase plate,
generating a highly non-symmetric wavefront, can be manufactured and
used in an imaging camera and telescope with AO system and produce a
scientifically useful degree of diffraction suppression.  The APP shape
can also be introduced using a deformable element with high spatial
frequency in upcoming AO systems to null out regions of interest, but
this requires a high degree of closed loop control and also limits the
bandwidth available for atmospheric correction. The APP wavefront has
regions of locally large wavefront tilt, and if introduced into an AO
loop, it can cause Shack-Hartmann wavefront sensors to become saturated.
By using a simple optic downstream of the wavefront sensing and
correction of the AO system, we eliminate the need for these schemes and
can implement, test and refine APP designs in short order.

The APP gives a significant detection advantage over direct imaging for
regions close to the star, typically over the range of $2-5\lambda/D$
for our initial measurements presented in this paper. By suppressing the
diffraction pattern and the quasi-static speckles that are tied to them,
we will then be limited by the noise characteristics of the uncorrected
AO halo, which are only just becoming significant in our data set.
Further observations will determine the effectiveness and sensitivity we
will reach with the APP in hour long exposures, and combined with a
refinement in the APP models, we will be able to detect point sources at
much smaller separations than is possible with direct imaging and
classical coronagraphs alone.

The suppression of diffraction limits the quasi-static speckle noise
that plagues the detection of point sources in imaging surveys for faint
companions at small angular separations. The APP is simple, efficient
and robust compared to other diffraction suppression methods, and can be
used in conjunction with other coronagraphic designs, making it a
potentially useful aid in the quest for direct imaging of extrasolar
planets.

\acknowledgments

This work was supported by NASA under grant NNG06GE06G and has been
encouraged by discussions with Nick Woolf.  We are grateful to Gregg
Davis at II-VI, and thank Elliot Solheid for drawing up the filter masks
and holders for the phase plate. We also thank our anonymous referee for
the many helpful comments and suggestions which helped improve this paper.
This work is based upon work supported by the National Aeronautics and
Space Administration through the NASA Astrobiology Institute under
Cooperative Agreement No.  CAN-02-OSS-02 issued through the Office of
Space Science.  The mathematical theory was developed in part under
grants from NASA, APRA04-0013-0056, and the NSF, AST-0138347.


{\it Facilities:} \facility{MMT}







\clearpage



\begin{deluxetable}{ccccc}
\tabletypesize{\scriptsize}
\rotate
\tablecaption{Position Angles, Separations, Differential Magnitudes and
Filters of \muh A\label{tbl-2}}

\tablewidth{0pt}
\tablehead{
\colhead{Date} & \colhead{P.A.} & \colhead{Separation} &
\colhead{$\Delta m$} & \colhead{Filter}
}
\startdata
2006.28070 & $214.55\pm0.20$ & $1.4907\pm0.0048$ & --- & M \\
2006.28072 & $214.69\pm0.18$ & $1.4847\pm0.0049$ & $4.99\pm0.06$ & M' \\
\enddata
\end{deluxetable}

\clearpage



\end{document}